\documentstyle[]{l-school}
\def\be{\begin{equation}}
\def\ee{\end{equation}}
\def\d0h{\hat \partial_0}
\def\Nb{\tilde N}
\def\bnab{\bar \nabla}

\begin{document}
\markboth{A. Abrahams and J. York}{3+1 General Relativity}
\setcounter{part}{1}
\title{3+1 General Relativity In Hyperbolic Form}
\author{A. M.  Abrahams and J. W. York Jr. }

\institute{Department of Physics and Astronomy\\
         University of North Carolina\\
         Chapel Hill, NC 27599-3255
}

\maketitle
\section{INTRODUCTION}

Numerical relativity represents the only currently
viable method for obtaining solutions to Einstein's 
equations for highly dynamical and strong field sources
of gravitational radiation.   The most astrophysically
interesting example is probably the final stages of 
binary inspiral and coalescence.  Partly motivated
by the prospect of observations with the next generation
of gravitational wave detectors, a multi-institutional 
``Grand Challenge'' effort\cite{Matz95} is underway in 
the US aimed at solving the full Einstein equations numerically
for coalescing black hole binaries.  In addition to
the tremendous computational resource demands of this
problem, it has been realized for some time that, unfortunately, the 
standard 3+1 formulation of Einstein's theory
as a Cauchy problem (cf. \cite{Yor79})
is somewhat deficient because of the
difficulty of imposing boundary conditions which maintain
numerical accuracy (and in some cases the physical
correctness) of the solution.  

The problem of boundary conditions 
is most dramatically evident in the study of black hole spacetimes.
Inside black holes spacelike slices either a) run into singularities
causing termination of simulations, b) freeze their evolution 
necessitating the commitment of more and more computational
resources to the astrophysically irrelevant black hole throat
as the simulation progresses.
An obvious solution to this problem is to excise the 
interior of the black hole from the computational domain.
Since it is impossible to identify the event-horizon dynamically
during the course of a simulation, a possible alternative is
to use the apparent horizon (which can be located on a single timeslice)
and always lies inside the true horizon (assuming cosmic censorship holds).

Numerical techniques based on the notion of causal 
differencing (cf. Seidel/Suen in this volume and
\cite{seidel_suen92}) have been proposed for
dealing with apparent horizon boundary conditions.  In
practice, it seems clear that the success of these
techniques is crucially dependent on the form of
the Einstein equations used\cite{bms95,A94}.
For spacetimes including gravitational radiation
a purely hyperbolic evolution system is imperative because
boundary conditions for the full set of constraint equations
are not available on the apparent horizon.  Furthermore, a purely
hyperbolic evolution scheme with ``simple'' characteristics, where the
only nonzero propagation speed is the speed of light,
enables one to ignore entirely the spacetime
inside the apparent horizon without concern that 
any unphysical (gauge) field quantities {\it should} be escaping
the horizon.

Outer boundary conditions for simulations on
spacelike slices of asymptotically
flat spacetimes are another important issue for the computation
of gravitational waveforms.  Since it is not 
feasible to simulate out to spatial infinity
where there is no radiation, it is important to have boundary
conditions that allow radiation to pass cleanly off the mesh.
If an outgoing boundary condition is applied to the wrong 
variable, spurious radiation is produced which can contaminate
the computed gravitational waveform.  Additionally, for some 
problems it is necessary to put the outer boundary at such a small radius
from the isolated source that backscatter of radiation off
curvature is significant.   This source of incoming radiation
then needs to be built into the outer boundary conditions.
The usual approach is to match the interior numerical
solution onto perturbation theory for the exterior
region\cite{ae88,ae90,ast95,ap96,AAERY96}. Work is also underway to allow the 
connection of interior Cauchy solutions to exterior numerical 
solutions on characteristic hypersurfaces\cite{winicour95}.
Both approaches benefit greatly from the use of a hyperbolic
evolution scheme with simple characteristics
for the interior solution.  Outer boundary values can be
assigned without the necessity of forming complicated 
admixtures of gauge and physical data.

In this paper we motivate and discuss a remarkable new hyperbolic formulation
of general relativity\cite{CBY95,AACBY95}
which may be thought of as a natural extension
of the usual 3+1 split of spacetime.  
This formulation preserves complete spatial covariance
by means of an arbitrary shift vector. The standard 3+1 treatment 
\cite{ADM,Yor79}, is gauge covariant in this sense but {\it not}
hyperbolic.  Naturally, our formulation does require a condition on
the time slicing to deal with the time-reparametrization invariance of 
the theory.  This is physically intuitive; for example, we believe that a 
complete understanding of the
generality of slicing conditions is a necessary first step towards
addressing the problem of time in quantum gravity.  
We expect many other applications of this formulation. Early
indications are that it will provide a powerful new approach to
perturbation theory and approximation schemes for general
relativity.

The plan of this paper is as follows.
First we will motivate the derivation of wave equations for general
relativity by considering the vastly simpler case of a scalar
wave and show how causal boundary conditions can be implemented.  We
then turn our attention to electromagnetism to demonstrate the
general procedure for gauge theories for producing wave equations.
We then take general relativity in 3+1 form and apply the
same method to obtain an explicitly hyperbolic formulation.  This
formulation is then written in first-order symmetric hyperbolic
form ideal for numerical implementation.  Finally we discuss perturbative
reductions of these equations and their use in outer boundary conditions
and radiation extraction.

\section{BOUNDARY CONDITIONS FOR THE SCALAR WAVE EQUATION}

Consider the simple scalar wave equation
in flat space:
\begin{equation}
\Box \psi = 0.
\end{equation}
Boundary conditions for this equation are straightforward to 
state and implement because the equation is manifestly hyperbolic.
Since the equation is linear, it is clear how to impose
outgoing wave boundary conditions at the edge of a numerical
domain.  It is also possible to employ inner "no boundary'' or
causal boundary conditions at  the edge of an expanding
null-surface (analogous to a black hole). 
Not surprisingly, since the equation is purely hyperbolic,
stable and accurate solutions can be obtained by merely ignoring
the causally disconnected region inside the null boundary.  
\begin{figure}
\centerline{\epsfig{file=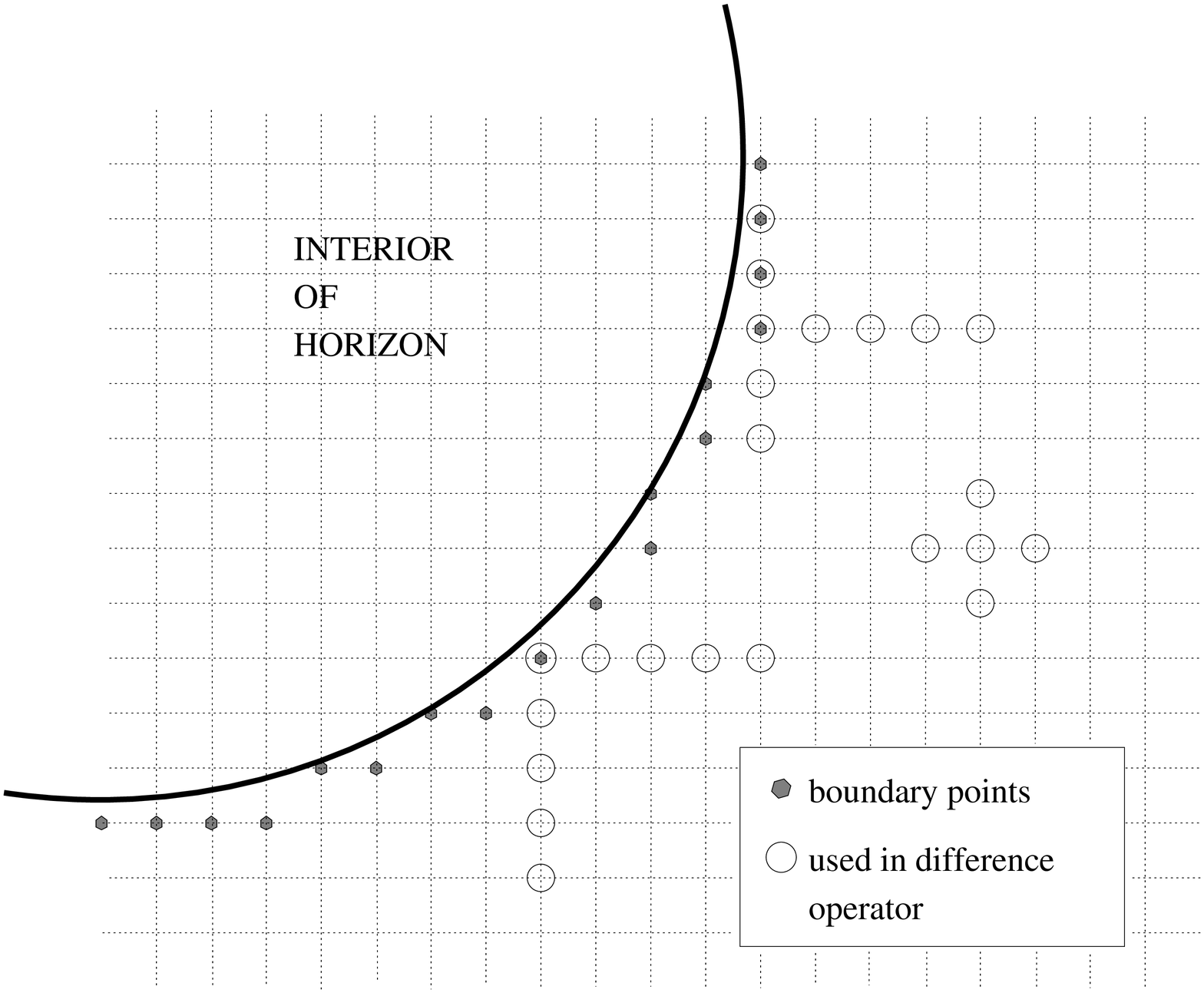,height=12cm,width=12cm}
}
\caption{Schematic diagram of numerical differencing
around a causal boundary. }    
\label{Fig1}
\end{figure}
 
We have performed tests of this notion using the simple scalar
wave equation (and straightforward nonlinear extensions) in 
2+1 dimensions and Cartesian coordinates:
\be
{\partial^2 \psi \over \partial t^2} = {\partial ^2 \psi \over \partial x^2}
			+ {\partial ^2 \psi \over \partial y^2}
\label{eq:scalarwave}
\ee
An arbitrary point in the grid is used as the origin of an expanding
spherical wavefront that itself is used as the inner boundary.  
Outgoing wave boundary
conditions are imposed at the rectangular outer boundary.  Equation
(\ref{eq:scalarwave}) is integrated using a predictor-corrector scheme.
The second spatial derivatives are computed using centered differences
except at the boundaries.  At every timestep, the expanding inner 
boundary is located, boundary points identified and special finite
difference operators constructed as shown in Figure~\ref{Fig1}.  Enough
points are used so that the 2nd-order error term can be set equal to the
2nd-order error in the centered difference scheme used in the 
mesh interior.  

We also simulate the effect of a shift-vector
or grid velocity by allowing the coordinate system to be time dependent.  This
is accomplished by a high-order interpolation step following each
time-step update.  Since this shift-vector can be larger than
the speed of propagation, it is possible for grid points to ``emerge''
from inside the horizon.  These points are filled with data from
outside using high-order extrapolation.
This code has been extensively tested with grid
velocities up to 5 times the propagation speed.  
It is stable, second-order convergent, 
and equal in accuracy to a comparison code 
using standard boundary conditions.
Nonlinear source terms seem to present no problems
for this scheme.  A similar, equally successful, algorithm 
has been developed in the context of a flux-conservative first-order version
of (\ref{eq:scalarwave}) and a Lax-Wendroff evolution scheme\cite{AL96}.

\section{DEVELOPMENT OF HYPERBOLIC SYSTEM}

In the previous section we have demonstrated
that hyperbolic wave equations
are very amenable to imposition of causal boundary conditions.  
Here we discuss the construction of analogous wave
equations for considerably more complicated 
(and gauge dependent) theories.

\subsection{Electromagnetism}

As a first example of a gauge theory, consider Maxwell's
equations in flat space, written here in 3+1 form\cite{smarr_york}.
The dynamical equations are
\begin{equation}
\partial_t A_i = -E_i - \nabla_i \phi 
\ee
\be
\partial_t E_i = -\nabla^j \nabla_j A_i + \nabla_i \nabla^j A_j - 4 \pi J_i
\ee
where $A_i$ is the vector potential, $E_i$ is the electric field,
$J_i$ is the current source,
$\nabla$ is a 3D flat space covariant derivative, and
$\phi$ is the gauge variable (scalar potential).  
These equations are supplemented by
the initial value constraint 
\begin{equation}
\nabla^j E_j = 4 \pi \rho
\label{eq:econ}
\end{equation}
with $\rho$ the charge density,
and a gauge condition on $A_i$ which requires the computation of $\phi$.
To produce a wave equation, one approach is to take a time derivative
of the $A_i$ evolution equation and substitute the $E_i$ evolution equation. 
To produce a D'Alembertian operator, it is necessary to apply, for
example, a transversality condition on $A_i$ which in turn 
imposes a radiation gauge condition on $\phi$: 
$\nabla^i A_i =0 \rightarrow \nabla^i \nabla_i \phi = -4 \pi \rho$
(using the continuity equation).
We have obtained a wave equation for $A_i$ in the ``Coulomb gauge.''
Alternatively, we could employ the Lorentz gauge:
$\partial_t \phi + \nabla^i A_i = \nabla^{\mu} A_{\mu} =0$
to obtain a wave equation for $A_\mu=(-\phi, A_i)$

An alternative, gauge-covariant, approach is to take a 
time-derivative of the evolution
equation for the {\it electric field}:
\begin{equation}
\partial^2_t E_i = \nabla_i \nabla^j (-E_j -\nabla_j \phi)
-\nabla^j \nabla_j (-E_i - \nabla_i \phi) - \partial_t J_i
\label{eq:dte}
\end{equation}
and use the constraint (\ref{eq:econ}) to eliminate the first
term yielding the wave equation
\begin{equation}
\Box E_i = 4 \pi  \nabla_i \rho - \partial_t J_i
.
\label{eq:wavee}
\end{equation} 
Interestingly, $A_i$ doesn't appear in (\ref{eq:wavee}); the dynamics
of electromagnetism have been cleanly separated from the
gauge-dependent evolution of the vector and scalar potentials.

\subsection{General Relativity}

Consider a globally hyperbolic manifold of topology 
$\Sigma\times R$ with metric $g_{\mu \nu}$.  A foliation of this spacetime is 
defined by a closed 1-form ${\bf\omega}=\nabla_\alpha t$
where $t$ is this coordinate time function and 
${\bf \omega}$ has normalization $||{\bf \omega}|| = -N^2$ with
$N$ the lapse function.
The four-dimensional line-element associated with
$g_{\mu \nu}$ may be decomposed in the general ADM\cite{ADM} form as
\begin{equation}
ds^2 = -N^2 dt^2 +g_{ij} (dx^i +\beta^i dt) (dx^j +\beta^j dt),
\ee
where $\beta^i$ is the shift vector.  It is convenient to
introduce the 
non-coordinate co-frame, 
\begin{equation}
\theta^0 = dt, \quad
\theta^i = dx^i+\beta^i dt
\end{equation}
with corresponding dual (convective) derivatives
\begin{equation}
\partial_0 = \partial/\partial t -\beta^i \partial/\partial x^i, \quad
\partial_i = \partial/\partial x^i. 
\end{equation}
In this non-coordinate basis the ADM metric takes the simple form:
\be
d s^2 = -N^2 (\theta^0)^2 + g_{ij} \theta^i \theta^j
.
\ee

Note that $[\partial_0,\partial_i]=(\partial_i \beta^k)\partial_k
=C_{0i}\vphantom{|}^{k}\partial_k$, where the
$C$'s are the structure functions of the co-frame, 
$d\theta^{\alpha}= -{1\over 2}C_{\beta\gamma}\vphantom{|}^{\alpha} 
\theta^{\beta}\wedge \theta^{\gamma}$. 

Instead of using the usual time-congruence 
$\partial/\partial t=\partial_0+\beta^k \partial_k$ which
follows the spatial coordinates to define
our evolution direction, we define a more
natural time derivative for evolution \cite{Yor79}
\begin{equation}
\hat\partial_0=\partial_0 +\beta^k \partial_k -{\cal L}_{\beta}
=\partial/\partial t -{\cal L}_{\beta},
\end{equation}
where ${\cal L}_{\beta}$ is the Lie derivative in a time slice $\Sigma$ along 
the shift vector.  In combination with the lapse as
$N^{-1}\hat\partial_0$, this is the derivative with respect to proper time 
along the normal to $\Sigma$, and it always lies inside the light cone, 
in contrast to $\partial/\partial t$.  In addition,  it has the useful
property that it commutes with the spatial coordinate derivatives, 
$[\hat\partial_0,\partial_i]=0$.  This time-derivative is particularly
appropriate to our form of the equations as we have subtracted
out the momentum constraints which, in the Hamiltonian formulation,
turn out to be generated by the shift-vector.
The dynamical variables in the standard 3+1 decomposition
are the 3-metric $g_{ij}$ and the extrinsic curvature of 
the slice $\Sigma$ as defined by the relation
\be
\d0h g_{ij} = -2N K_{ij} .
\label{eq:gev}
\ee

In four dimensions, we can write Einstein's equation as
\be
R_{\mu\nu}=\kappa (T_{\mu\nu}
-{1\over 2}g_{\mu\nu} T^{\lambda}\vphantom{|}_{\lambda}).
\label{eq:ee}
\ee
Using the moving basis defined above, (\ref{eq:ee}) can be
split up into constraints and evolution equations.
The time-time part of the Einstein equation leads to 
the Hamiltonian constraint
\be
R^0_0 - R^i_i  = -\bar R - H^2 + K_{ij} K^{ij}
\label{eq:hamcon}
\ee
where $H \equiv K^i_i$ and $\bar R$ is the trace of the
3-dimensional Ricci tensor $\bar R_{ij}$ (barred quantities
are always spatial in our notation with indices running from
1-3).  
The time-space parts of Einstein's equation
yield the momentum constraints:
\be
 N^{-1} R_{0i} = \bar \nabla_j K^j_i - \bar \nabla_i H
.
\label{eq:momcon}
\ee 
The purely spatial parts of Einstein's equation give us
the evolution of the extrinsic curvature:
\be
R_{ij}= -{1 \over N} \d0h K_{ij} + H K_{ij} - 2 K_{i l}K^l_j
-{1 \over N} \bnab_i \bnab_j N + \bar R_{ij}
.
\label{eq:kev}
\ee
The standard 3+1 formulation of general relativity consists
of the evolution equations (\ref{eq:gev}) and (\ref{eq:kev})
with initial data $(g_{ij}, K_{ij})$ satisfying the constraints
(\ref{eq:hamcon}) and (\ref{eq:momcon}).  These equations
are supplemented by equations for the sources (if any)
and for the kinematical variables
$\beta^i$ and $N$.  The "slicing'' equation for $N$
often determined by a condition on $H$ 
using the trace of (\ref{eq:kev}).

To derive a wave equation for general relativity, one could,
for example, follow the classic method of eliminating from
$R_{\mu \nu}$ the unwanted second derivatives of $g_{\mu \nu}$
by using the full spacetime harmonic condition $\Gamma^\mu = 0$
\cite{CB52,CBY79,FiM}.  One would then obtain a non-geometric D'Alembertian 
$g^{\alpha \beta} \partial_\alpha \partial_\beta g_{\mu \nu}$.
This procedure is analogous to using the Lorentz gauge in Maxwell's
theory.  

Instead, we shall follow the spatially gauge-covariant analog
of the procedure that led to (\ref{eq:dte}) and (\ref{eq:wavee})
for Maxwell's equations.  The spatial metric  $g_{ij}$ is
analogous to $A_i$, the shift $\beta^k$ to $\phi$,
and the extrinsic curvature $K_{ij}$ of $\Sigma$ to $E_i$.
The lapse, on the other hand, is a quantity found only
in time-reparametrization invariant theories.  We take a time
derivative of the equations of motion and subtract spatial
gradients of the momentum constraints, thus obtaining
a new quantity $\Omega_{ij}$
\begin{equation}
\label{omega}
\hat\partial_0 R_{ij} -\bar\nabla_{i} R_{0j}-\bar\nabla_{j}
R_{i0}=\Omega_{ij}.
\end{equation}
In terms of the dynamical gravitational variables, $\Omega_{ij}$
may be expressed as 
\begin{equation}
\label{boxK}
\Omega_{ij}=N\Box K_{ij} + J_{ij} +S_{ij},
\end{equation}
where $\Box=-N^{-1}\hat\partial_0 N^{-1}\hat\partial_0 
+\bar\nabla^{k} \bar\nabla_{k}$ is the physical wave operator for
arbitrary $\beta^k$.  It is constructed from 
second proper time-derivatives and
second covariant spatial-derivatives.
The source term is given by
\begin{eqnarray}
\label{Jij}
J_{ij}&=& \hat\partial_0 (H K_{ij} - 2 K_{i}\vphantom{|}^{k} K_{jk})
 +(N^{-2}\hat\partial_0 N+ H)\bar\nabla_i \bar\nabla_j N \nonumber \\
&&\hspace{-0.75cm}
-2N^{-1}(\bar\nabla_k N) \bar\nabla_{(i}(N K^{k}\vphantom{|}_{j)})
+3 (\bar\nabla^k N) \bar\nabla_k K_{ij} \\
&&\hspace{-0.75cm} +N^{-1}K_{ij} \bar\nabla^k (N\bar\nabla_k N)
-2 \bar\nabla_{(i}(K_{j)}\vphantom{|}^{k}\bar\nabla_k N)
+N^{-1} H \bar\nabla_i\bar\nabla_j N^2
\nonumber \\
&&\hspace{-0.75cm}
+2 N^{-1}(\bar\nabla_{(i} H)(\bar\nabla_{j)}N^2 )
 -2N K^{k}\vphantom{|}_{(i}\bar R_{j)k}
-2N \bar R_{kijm}K^{km}. \nonumber
\end{eqnarray}
and contains no second-derivatives of the extrinsic curvature.

The slicing term $S_{ij}$ is given by 
\begin{equation}
S_{ij}=-N^{-1}\bar\nabla_i \bar\nabla_j(\hat\partial_0 N +N^2 H).
\end{equation} 
For $\Omega_{ij}$ to produce a wave equation, $S_{ij}$  must be equal 
to a functional involving fewer than second derivatives of
$K_{ij}$.  The most obvious way to guarantee this is
to use the harmonic condition (cf. \cite{CBR} when $\beta^k=0$)
\begin{equation}
\label{d0N}
\hat\partial_0 N +N^2 H=0.
\end{equation}
(This can easily be generalized by adding an ordinary well behaved function 
$f(t,x)$ to the right hand side.  The slicing generality compatible
with hyperbolic evolution schemes is the subject of Ref.~\cite{AACBY96}.) 
Imposing (\ref{d0N}) for all time amounts to imposing an equation of motion 
for $N$. The shift is completely arbitrary: it can be given as
a function of space and time or solved for on each timestep 
based on some condition on the evolved variables.
This ``System I'' is purely hyperbolic.  Initial data consists
of 3-metric and extrinsic curvature satisfying the constraints
and proper time-derivative of the extrinsic curvature satisfying
(\ref{eq:kev}).  The initial value for the lapse must also be specified.

An alternative to the harmonic condition is to
specify the trace of the extrinsic curvature as a known 
function for all time $H=h(x,t)$.  This also eliminates the second
derivatives of unknown functions in $S_{ij}$ and provides a 
time-dependent 
elliptic equation,
\begin{equation}
\hat \partial_0 h= - \bar\nabla^k\bar\nabla_k N
+ N (\bar R + H^2 -g^{ij}R_{ij}),
\label{eq:d0H}
\end{equation}
to solve for $N$ on each timestep.
The shift vector
is still freely specifiable.
This ``System II'' is mixed hyperbolic/elliptic. 
It is possible to prove\cite{CBY95} using the doubly
contracted Bianchi identity, that Systems I and II 
are completely equivalent to Einstein's theory.
These systems are summarized in Table \ref{Table1}.
\begin{table}
\caption{Possible evolution systems}
\begin{tabular}{ l  c  c}
\hline 
System (type) & Equations & Initial Data \\
\hline 
System I&$\d0h g_{ij}= -2N K_{ij}$ & $g_{ij}$ \\ 
(hyperbolic)&~$N \Box K_{ij}= \Omega_{ij}-J_{ij}+\bnab_i \bnab_j f(t,x)$~~   
&~$K_{ij}$, $\d0h K_{ij}$ \\
&$\d0h N + N^2 H = f(t,x)$&$N$  \\
\hline 
\hline
System II&$\d0h g_{ij}= -2N K_{ij}$&$g_{ij}$ \\
(Mixed hyperbolic/&~$N \Box K_{ij}= \Omega_{ij}-J_{ij}-S_{ij}$~~& 
~$K_{ij}$, $\d0h K_{ij}$ \\
elliptic)&
$\hat \partial_0 h= - \bar\nabla^k\bar\nabla_k N
+ N (\bar R + H^2 -g^{ij}R_{ij})$& $H=h(t,x)$ \\
\hline
\end{tabular}
\label{Table1}
\end{table}

It is easily seen that for first-order perturbations of
static backgrounds, the evolution equation for 
the 3-metric (\ref{eq:gev}) and the wave equation for the extrinsic
curvature (\ref{boxK}) become completely decoupled.
We will explore
this idea further in the section on perturbative reduction.
This situation is analogous to the separation of the equation for
$A_i$ from the wave equation for $E_i$ we saw in 
linear electromagnetism.  It is also consistent with 
physical intuition about the separation of transverse 
wave motion from longitudinal fields in general relativity.
Locally, the 3-metric provides a background on top of 
which the extrinsic curvature propagates. 

\subsection{First-order hyperbolic form for System I}

In order to reduce System I to hyperbolic form 
it is necessary to define some new variables.
(Here we 
restrict ourselves to the vacuum case and the simple
harmonic slicing condition $f(t,x)=0$.)
We introduce $a_i=N^{-1}\bar\nabla_i N$---the acceleration of the local 
Eulerian observers (those at rest in the time slices)---its derivatives 
$a_{0i}=N^{-1}\hat\partial_0 a_i$ and
$a_{ji}=\bar\nabla_j a_i=a_{ij}$, as well as time and
space derivatives of
the extrinsic curvature 
\begin{equation}
\label{eq:d0K}
\hat\partial_0 K_{ij}=N L_{ij}
\end{equation}
and
$M_{kij}=\bar\nabla_k K_{ij}$.
System I can now be cast in
flux-conservative first-order symmetric hyperbolic form\cite{CBY95,AACBY95}.
The 49 unknowns of the first-order system are $g_{ij}$, $N$, $K_{ij}$
$L_{ij}$, $M_{kij}$, $a_i$, $a_{ji}$ and $a_{0i}$, and the equations 
are  (\ref{eq:gev}), (\ref{d0N}), (\ref{eq:kev}),
(\ref{eq:d0K}, and
\begin{equation}
\hat\partial_0 L_{ij} - N \bar\nabla^{k} M_{kij} = J_{ij},
\end{equation}
\begin{eqnarray}
\hat\partial_0 M_{kij} -N \bar\nabla_{k} L_{ij} &=&
N[ a_k L_{ij} + 2M_{k(i}\vphantom{|}^{m} K_{j)m}\\
&&\hspace{-2.5cm} + 2K_{m(i} M_{j)k}\vphantom{|}^{m} 
-2K_{m(i} M^{m}\vphantom{|}_{j)k}  \nonumber\\ 
&&\hspace{-2.5cm} + 2K_{m(i} ( K^{m}\vphantom{|}_{j)}a_k 
+ a_{j)} K^{m}\vphantom{|}_{k} -a^{m} K_{j)k})] , \nonumber
\end{eqnarray}
\begin{equation}
\hat\partial_0 a_i =-N(H a_i + M_{ik}\vphantom{|}^{k}),
\end{equation}
\begin{eqnarray}
\hat\partial_0 a_{ji} -N \bar\nabla_{j} a_{0i}&=& Na_{k} 
[2 M_{(ij)}\vphantom{|}^{k} -M^{k}\vphantom{|}_{ij} \\ 
&&\hspace{-2cm}+2 a_{(i} K_{j)}\vphantom{|}^{k} - a^{k} K_{ij}]
 + N a_j a_{0i}, \nonumber
\end{eqnarray}
\begin{eqnarray}
\hat\partial_0 a_{0i} - N\bar\nabla^{k} a_{ki} &=&
N [-\bar R^{k}\vphantom{|}_{i} a_{k} +a_{i}( H^2 -2 K_{kl}K^{kl} \\
&&\hspace{-3cm}+2 a^k a_k +2 a^{k}\vphantom{|}_{k}) 
+2 a_k a^{k}\vphantom{|}_{i} + H M_{ik}\vphantom{|}^{k}
-2K^{kl} M_{ikl}],  \nonumber
\end{eqnarray}
where $J_{ij}$ is computed using (\ref{Jij}). 
To reduce this system to completely first-order form, 
the 3-dimensional Riemann curvature appearing in $J_{ij}$ 
is expressed in terms of the 3-dimensional Ricci curvature using
the identity
\begin{equation}
\label{barR}
\bar R_{mijk}= 2 g_{m[j}\bar R_{k]i} + 2g_{i[k}\bar R_{j]m}+
\bar R g_{m[k} g_{j]i}.
\end{equation}
The 3D Ricci tensor is then eliminated using (\ref{eq:kev}) rewritten
in terms of first-order variables:
\begin{equation}
\bar R_{ij}= R_{ij} +L_{ij} -H K_{ij} +2 K_{ik}K^{k}\vphantom{|}_j + 
a_i a_j+ a_{ji}.
\end{equation}
The 4D Ricci tensor is computed from sources using (\ref{eq:ee}).
To demonstrate strictly first-order form it
is also necessary to make Christoffel symbols part of the
system by introducing a background metric as in Ref.~\cite{CBY95}.
We stress again that the shift vector is not one of the unknown fields;
the form of the equations is completely independent of $\beta^k$.

With this form of the equations it is possible to read off
the characteristic speeds of the different fields and
verify one's physical expectations about the propagating
degrees of freedom.
Since there are no spatial derivatives on the left-hand-side of
their evolution equations, we see that $g_{ij}$, $K_{ij}$, 
$N$, and $a_i$ all propagate with zero
speed with respect to the Eulerian observers: they glide up the
the normal to the foliation driven by the dynamical sources. 
Only the derivatives of the extrinsic curvature ($L_{ij}, M_{ijk}$)
and the derivatives of the acceleration ($a_{oi}$, $a_{ij}$)
propagate with the speed of light.  
These quantities all appear in the components of
the spacetime Riemann tensor and thus represent tidal fields.
 
Along with A. Anderson (UNC) we have performed numerical tests of this
form of the equations on the simple dynamical problem of
even and odd-parity cylindrical waves.  A Lax-Wendroff
scheme is easily coded for the 27 equations necessary to
describe this system with complete spatial gauge
freedom.  We find that a stable and accurate evolution can 
be computed which is comparable with that obtained by solving the usual
3+1 equations in fully harmonic coordinates (see \cite{piran85}
for this version of the equations).

\section{PERTURBATIVE REDUCTION AND OUTER BOUNDARY CONDITIONS}

Here we sketch the reduction of System I
for perturbations of the Schwarzschild metric.   Full details
are given in Ref.~\cite{AAERY96}. This reduction has
helped us elucidate many aspects of the full theory and provides
new insight into the nature of gauge-invariant perturbation
theory.  It also allows us to define a framework for 
both radiation extraction and outer boundary conditions
based on Schwarzschild perturbation theory.  Such a framework
can be used in conjunction with numerical simulations.

\subsection{First-order perturbation theory of Schwarzschild}

For first order perturbations of static Schwarzschild, we
make the following decomposition:
\begin{eqnarray}
g_{ij} &=& \tilde g_{ij} + h_{ij} \\
K_{ij} &=& 0 + \kappa_{ij} \\
N &=& \tilde N + \alpha \\
\beta^i &=& 0 + v^i.
\end{eqnarray}
Tildes denote background values.
The background metric and lapse take their standard static
Schwarzschild values and the background extrinsic curvature and
shift are zero.  Unless otherwise noted, covariant derivatives
are with respect to the background metric.

The evolution of the 3-metric (\ref{eq:gev}) reduces to
\begin{eqnarray}
\partial_t \tilde g_{ij} &=& 0 \\
\partial_t h_{ij} &=& - 2 {\tilde N} \kappa_{ij} + 2 \bnab_{(i} v_{j)}
.
\label{eq:dthsch}
\end{eqnarray}
Notice that this equation is entirely gauge dependent: the arbitrary
choice of shift $v^i$ translates into arbitrary distortion of
metric perturbations.  
The harmonic condition for the lapse splits into:
\begin{eqnarray}
\partial_t \Nb &=& 0 \\
\partial_t \alpha &=& v^i \partial_i \Nb - \Nb^2 \kappa
.
\label{eq:pertlapse}
\end{eqnarray}

The wave equation for the extrinsic curvature (\ref{boxK})
reduces to
\begin{eqnarray}\label{eq:BoxKsch}
{1 \over \Nb} \partial^2_t \kappa_{ij}- \Nb \bnab^k \bnab_k \kappa_{ij}=
  -4 \bnab_{(i}\kappa^k_{j)} \bnab_k \Nb
 + \Nb^{-1} \kappa_{ij} 
\bnab^k \Nb \bnab_k \Nb + 3 \bnab^k \Nb \bnab_k \kappa_{ij}
    \nonumber \\
+\kappa_{ij} \bnab^k\bnab_k 
\Nb -2 \kappa^k\mathstrut_{(i}\bnab_{j)}\bnab_k \Nb
 - 2{\Nb}^{-1} \kappa^k\mathstrut_{(i}\bnab_{j)}\Nb \bnab_k \Nb
+2\kappa \bnab_i \bnab_j \Nb
    \nonumber \\
+4 \partial_{(i}\kappa \partial_{j)} \Nb +
2\Nb^{-1} \kappa \bnab_i\Nb \bnab_j \Nb
-2\Nb {\tilde R}_{k(i}\kappa^k\mathstrut_{j)}
- 2\Nb {\tilde R}_{kijm} \kappa^{km}.
\end{eqnarray}
Here $\kappa=K^i\mathstrut_i$ and $\tilde R_{ij}$ and
$\tilde R_{ijkl}$ are background, spatial Ricci and Riemann
tensors respectively.
This equation is entirely decoupled from the evolution
of the 3-metric perturbation (\ref{eq:dthsch}).  
It could be directly evolved in an exterior region as a perturbative
version of the full evolution equations.  (Note that for
flat space, $\Nb = 1$, (\ref{eq:BoxKsch}) reduces to $\Box \kappa_{ij} =0$
which has radiative solutions corresponding to first time-derivatives
of the usual transverse-traceless metric perturbations for gravitational
waves.) 

Alternatively, one can perform a decomposition of (\ref{eq:BoxKsch})
in terms of tensor spherical harmonics and produce scalar
wave equations for the different $\ell, m$ mode combinations.
Here, for simplicity, we restrict attention to the slicing independent
odd-parity perturbations.
For odd-parity perturbations, $\kappa_{ij}$ is decomposed with
two tensor spherical harmonics and two amplitude functions.  The
component $\kappa_{r \phi}$ is expressed  in terms of
the amplitude function $a_\times$ and angular functions as 
\be
\kappa_{r \phi} = 
 a_\times (t,r) \sin \theta \partial_\theta Y_{\ell m} 
.
\ee
Taking the $r$-$\phi$ component of (\ref{eq:BoxKsch})
and utilizing the $\phi$ component of the momentum constraint,
a 1-dimensional scalar wave equation purely in terms of the amplitude
function $a_\times$ is formed:
\begin{eqnarray}
\left[\partial^2_t -(1-2M/r)^2\partial^2_r -(2/r)(1-2M/r)\partial_r
-2M/r^3+3M^2/r^4
\right.
\nonumber \\
\left. +(1-2M/r)(\ell(\ell+1)/r^2-6M/r^3) \right]a_\times(t,r)=0.
\label{eq:oddscal}
\end{eqnarray}
To form the standard Regge-Wheeler equation for odd-parity 
perturbations of Schwarzschild (cf. \cite{moncrief})
one takes a time-derivative of this equation using
\be
\partial_t \kappa_{r \phi} = - \bnab_r \bnab_\phi  \Nb+\Nb \bar R_{r \phi}
\ee
which is the perturbative reduction of (\ref{eq:kev}) for odd-parity
perturbations.  Here the covariant derivatives are with respect
to the perturbed background and the 3D Ricci tensor is computed
from the perturbed metric.  The variable $\partial_t a_\times$
satisfies the usual Regge-Wheeker equation.  
We note that no work has been required to construct spatial
gauge invariants.  These come ``for free'' in our spatially
covariant wave-equation. 

The situation for even parity perturbations is somewhat more
involved because the lapse perturbations (\ref{eq:pertlapse})
couple into
our wave equation via
the trace of $\kappa_{ij}$ and the harmonic slicing condition
at the extrinsic curvature level.  
The same basic procedure
holds, however.  Using a tensor spherical-harmonic decomposition
and the radial component of the momentum constraint, coupled
1D wave equations are formed for projections of
$\kappa_{rr}$ and of $\kappa$. 
Connection to the Zerilli equation can be made by
taking a time-derivative of these equations.
The usual gauge invariant perturbation equations
for Schwarzschild spacetime\cite{moncrief} are seen,
not surprisingly, to represent curvature evolution.

\subsection{Radiation extraction and outer boundary conditions}

Perturbation theory has proven to be a powerful tool for
extracting physical information from numerically generated
spacetimes (cf. \cite{ae88,ae90,ast95,ap96}).  The basic scheme is
to match the full nonlinear interior solution to perturbation theory
along the timelike cylinder representing the
boundary of the computational domain.
This idea is illustrated in Figure 2.
\begin{figure}
\centerline{\epsfig{file=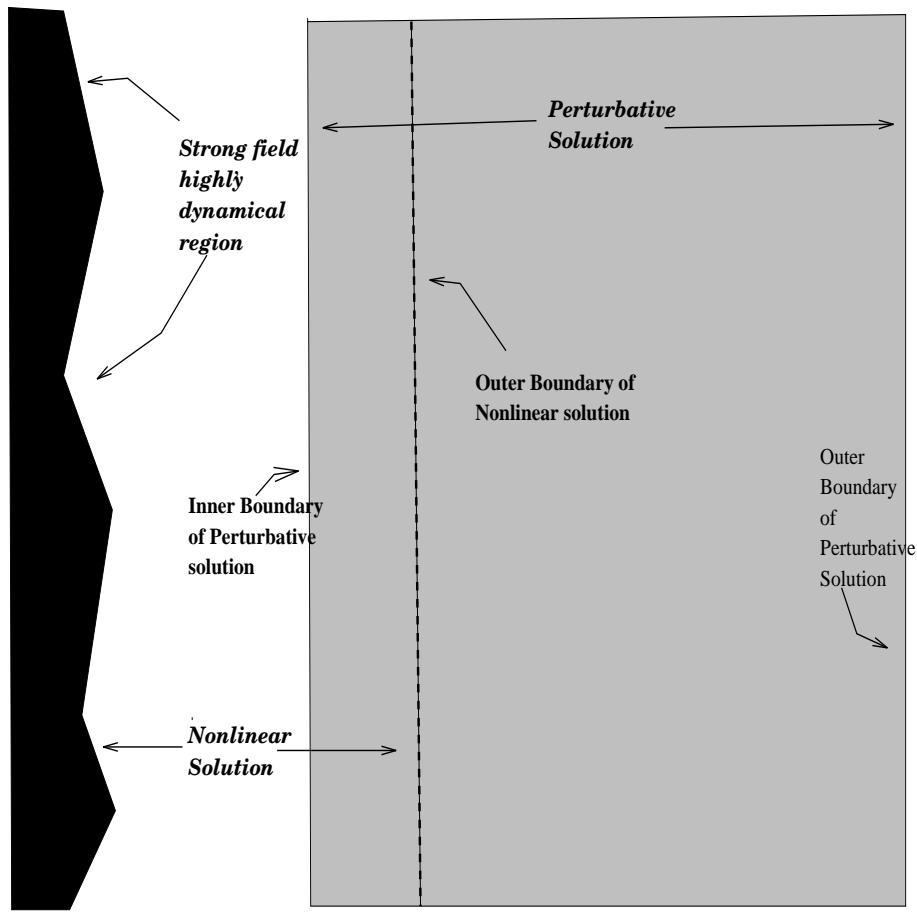,height=12cm,width=12cm}
}
\caption{Schematic diagram of a simulation with a solution 
to the full Einstein equations in the interior matched
onto a perturbation
theory solution in the exterior (shaded region).  
During the course of the evolution, boundary data for the perturbative
evolution is read off from the nonlinear solution at the 
inner boundary.  Data from the perturbative solution is used
in turn to construct outer boundary data for the nonlinear
simulation.  Approximate asymptotic waveforms are read off
the perturbative solution at large radius.
}    
\label{Fig2}
\end{figure}

The main steps in this scheme are to 
1) construct perturbatively gauge-invariant quantities from evolved code
variables, 2) propagate these gauge-invariants to large
radius to remove near-zone effects, 3) use information from step 2
to construct code variables at the edge of the mesh, thus providing
outer boundary conditions.
The construction of gauge-invariants makes it possible to use the
same extraction procedure in conjunction with numerical simulations
using different choices of spatial gauges.
This follows closely the conceptual picture 
for the calculation of gravitational radiation from isolated 
sources laid out by Thorne \cite{thorne80}. The calculation of the 
strong field and dynamical source is performed by the 
numerical simulation.  The overlap/matching region is in the 
nondynamical near zone region (within a typical wavelength of the source). 
The goal is to compute waveforms in the wave zone beyond which
the geometric optics approximation can be used to propagate the
waves.  In the procedure shown in Fig.~\ref{Fig2}, the waveform
is read off the perturbative variables at
the outer boundary of the exterior evolution. 
Effects of backscatter
off background curvature between the outer boundary of the 
interior nonlinear solution and the outer boundary of the
exterior perturbative solution have been taken into
account in both the waveform and the boundary conditions
imposed on the interior solution.

As should be clear from the discussion in the previous section,
the new hyperbolic formulation elucidates the process of
attaching the standard 3+1 variables $g_{ij}$,
$K_{ij}$ etc. onto perturbation theory.  This subject
is explored fully both for weak-field and Schwarzschild
perturbation theory in Ref.\cite{AAERY96}.
In the weak field case, the exterior perturbative evolution
can be done analytically.  For Schwarzschild
perturbations this requires a straightforward numerical
integration using the same coordinate time steps as the
interior evolution.  Boundary data for the
exterior equations is computed via multipolar projections of
the components of $K_{ij}$ and $L_{ij}$. 
The Schwarzschild mass is found from the ADM surface
integral performed near the edge of the interior mesh.  

To produce boundary data for the interior simulation, 
the components of $K_{ij}$ and $L_{ij}$ are reformed from the 
perturbative variables using the momentum constraint equations.  
For System I, the lapse is determined by the harmonic slicing
condition which ties it to the trace of the extrinsic curvature.
So the lapse at the outer boundary is set directly from the
exterior evolution. (If an elliptic slicing condition is
used, then the lapse at the outer boundary will be known
given boundary values for the extrinsic curvature and an
imposed condition on (\ref{eq:d0H}).)
The boundary condition on the 3-metric is
determined using (\ref{eq:gev}), the known boundary
values for the extrinsic curvature and the lapse,
and the chosen boundary condition
on the shift vector components.

As mentioned in the previous section, an alternative 
procedure for the exterior evolution is to simply integrate 
Eq. (\ref{eq:BoxKsch}) in the exterior on a 3D finite 
difference grid.  This can be accomplished either using
a Cauchy or characteristic formulation of the equation
and a spherical polar topology numerical mesh for computational
efficiency.   Since the coordinate singularity at $r=0$ will 
not be part of the evolution domain, the usual difficulties with
numerical instabilities will be avoided.  In addition, it
will be sufficient to perform adaptive mesh refinement, if desired,
in only the radial direction.
Imposition of boundary values is trivial for the extrinsic curvature
and proceeds exactly as above for the other variables.

\ack{
A.A. would like to thank the organizers of the Les Houches workshop
for giving him the opportunity to present this
work in such a beautiful location.  The authors have benefited 
greatly from their collaboration with Arlen Anderson and 
Yvonne Choquet-Bruhat on the development of the 
hyperbolic systems and we thank them for their essential contributions.
We also thank Mark Rupright for 
collaborating on the perturbative reduction of the equations.
The research described here was supported by 
National Science Foundation grants PHY-9413207 and PHY 93-18152/ASC 
93-18152 (ARPA supplemented). 
}


\begin{thebibliography}{100}

\bibitem{Matz95}  proceedings of November 1995
Grand Challenge Alliance workshop may be obtained by
contacting R. Matzner at U. Texas, Austin.

\bibitem{Yor79}
J.W. York,
in {\it Sources of Gravitational Radiation}, edited by L. Smarr,
(Cambridge Univ. Press: Cambridge, 1979).

\bibitem{seidel_suen92}
Seidel, E. and Suen, W.-M.,
\review Phys. Rev. Lett.,  69, 1992, 1845.

\bibitem{bms95}Bona, C., Mass\'{o} , J., and Stela, J.,
\review Phys. Rev. D, 51, 1995, 1639.

\bibitem{A94} Abrahams, A.,
in unpublished proceedings of November 1994 Grand Challenge
Alliance meeting, edited by E. Seidel, NCSA. 

\bibitem{ae88}
Abrahams, A.M. and Evans, C.R.,
\review Phys. Rev. D, 37, 1988, 317.

\bibitem{ae90}
Abrahams, A.M. and Evans, C.R.,
\review Phys. Rev. D, 42, 1990, 2585.

\bibitem{ast95} Abrahams, A.M., Shapiro, S.L.
and Teukolsky, S. A.
\review Phys. Rev. D, 51, 1995, 4295.

\bibitem{ap96}
Abrahams, A.~M. and Price, R.~H.,
{\it Phys. Rev. D}, in press (1996).

\bibitem{AAERY96}
Abrahams, A., Anderson, A., Evans, C.,
Rupright, M. and York, J.W. in preparation.

\bibitem{winicour95} Winicour, J. in Ref. 1.

\bibitem{CBY95}
Choquet-Bruhat, Y. and  York, J.W. C.~R.~Acad.~Sci. Paris, {\bf 321}, 
1089 (1995).

\bibitem{AACBY95} Abrahams, A., Anderson, A., Choquet-Bruhat, Y.,
and York, J.W., Phys. Rev. Lett. {\bf 75}, 3377 (1995).

\bibitem{ADM}
R. Arnowitt, S. Deser and C.W. Misner,
in {\it Gravitation}, edited by L. Witten,
(Wiley: New York, 1962).

\bibitem{AL96}
Abrahams, A. M. and Lenaghan, J., in preparation (1996).

\bibitem{smarr_york} 
Smarr, L.L., York, J.W.
\review Phys. Rev. D, 17, 1978, 1945.

\bibitem{CB52} Choquet-Bruhat, Y.
\review Acta. Math.,  88, 1952, 141.

\bibitem{CBY79}
Choquet-Bruhat, Y. and J.W. York,
in {\it General Relativity and Gravitation}, edited by A. Held
(Plenum: New York, 1979).
	 
\bibitem{FiM} Fischer, A. and Marsden, J.
\review Comm. Math. Phys.,  28, 1972, 1.

\bibitem{CBR}
Choquet-Bruhat, Y. and Ruggeri, T. 
\review Comm. Math. Phys., 89, 1983, 269.

\bibitem{AACBY96}
Abrahams, A., Anderson, A., Choquet-Bruhat, Y. and York, J.W. in preparation.

\bibitem{piran85} Piran, T., Safier, P.N., Stark, R.F.,
\review Phys. Rev. D., 32, 1985, 3101.

\bibitem{moncrief}
Moncrief, V.
\review  Ann. Phys. (NY), 88, 1974, 323.

\bibitem{thorne80}Thorne, K.~S.,
\review Rev.~Mod.~Phys., 52, 1980, 299.
 
 
\end{thebibliography}
\end{document}